\documentclass{article}

\usepackage{hyperref}
\usepackage{color}
\usepackage{graphicx}
\usepackage{amstext,amsmath,amssymb,amsfonts,bbm,
amsthm,MnSymbol}
\usepackage{fancyhdr}
\usepackage{subfigure}
\usepackage{float}

\usepackage{vmargin}

\setpapersize{A4}
\setmargins{2.5cm}       
{1.5cm}                        
{16.5cm}                      
{23.42cm}                    
{10pt}                           
{1cm}                           
{0pt}                             
{2cm}                           

\begin{document}


\title{Light Propagation in the vicinity of the ModMax black hole.}

\author{E.Guzman-Herrera\footnote{eguzman@fis.cinvestav.mx} and 
N. Breton\footnote{nora@fis.cinvestav.mx} \vspace{1cm}\\
Physics Department \\
Centro de Investigaci\'{o}n y de Estudios Avanzados del I. P. N. \\ 
Apdo. 14-740, CDMX, M\'{e}xico}

\maketitle

\begin{abstract}
ModMax is a nonlinear electrodynamics theory with the same symmetries as Maxwell electrodynamics. Static spherically symmetric solutions have been derived by coupling ModMax electrodynamics with the Einstein equations, which can represent a black hole. In this paper, we analyze light propagation in the vicinity of the ModMax black hole. We determine birefringence, light trajectories, deflection, redshifts, as well as the shadow of the black hole using the effective or optical metric to determine the optical paths of light;  comparison is done with the corresponding effects in the neighborhood of the Reissner-Nordstrom black hole, that is the solution to the Einstein-Maxwell equations.
\end{abstract}
\vspace{1cm}


\newpage
\section{Introduction}

At a macroscopic level and even at an atomic level, the linear superposition of electromagnetic fields is confirmed experimentally with a precision of less than $1\%$. Nonetheless, at a subatomic level, there are deviations from the linear superposition, this is due to the proximity between charged particles and the high intensity of their fields, then the Maxwell electromagnetic theory presents singularities. 
In the presence of intense electromagnetic fields (EM), light propagation can be modeled as if moving through a dispersive medium \cite{Novello2000}. When the electromagnetic field strength approaches the critical electric field $E_{cr}=m_{e}^2 c^3 / (e \hbar) \approx 10^{18}V/m$ or 
the critical magnetic field $B_{cr}=10^9$T, the effect of external fields on vacuum quantum properties becomes significant. These effects can be described phenomenally by classical theories characterized by Lagrangians that depend nonlinearly on the two electromagnetic invariants. 

Two of these theories stand out: the Euler-Heisenberg (EH) and the Born-Infeld (BI).  The BI theory solves the divergence of the self-energy of point particles however, this theory is not conformally invariant.
The EH nonlinear electrodynamics (NLED) was derived from the QED Lagrangian in the tree-level approximation
and phenomenologically describes the effect of vacuum polarization; however, the EH NLED theory is neither conformal invariant nor dual invariant. 

The question arose if there is any nonlinear electrodynamics fulfilling the same symmetries as Maxwell electrodynamics, namely,  the four-dimensional conformal symmetry and the electric-magnetic duality. The answer is yes. Recently in \cite{Bandos2020} was proposed a NLED theory with the two symmetries;  it is characterized by a dimensionless parameter $\gamma$ and it reduces to Maxwell theory if $\gamma=0$. This theory, known as Modified Maxwell (ModMax) NLED, has stimulated research in several aspects, from classical solutions \cite{Banerjee2022, Neves2023} to super symmetric analysis \cite{Escobar2022, Bandos2021p, BandosSusy, Barrientos2022, Bordo2021, Bokulic2021, Pantig2022, Babaei2022}. 

On the other hand, at strengths approaching the critical electromagnetic fields, self-field interactions induce modifications in light trajectories, in a way similar to light propagating in a dispersive media, for instance in anisotropic materials; alternatively, it can be modeled as light propagating through a curved spacetime. Deviations from the trajectories in vacuum are described in NLED by the null trajectories of an effective metric (also known as the optical metric) that is derived from the analysis of the propagation of electromagnetic field discontinuities  \cite{Novello2000, Pleban, Novello2000b}. 

Coupling the ModMax electrodynamics with Einstein gravity, for a static and spherically symmetric (SSS) metric, black hole (BH) solutions were found.
In the present paper, we focus on light propagation in the neighborhood of a ModMax BH, determining light trajectories as the null geodesics of the effective metric. 

This paper is organized as follows:  In section \ref{sect:EffectiveMetric} we present the expression for the effective or optical metric induced by a general nonlinear Lagrangian in an arbitrary background metric.
In section \ref{sect:ModMax} we present the ModMax NLED Lagrangian.  In section \ref{sect:ModMaxBH} we present the main features of the SSS ModMax-Einstein BH as the background metric. In section \ref{sect:Birefringence}, using the effective metric we analyze the birefringence in the dyonic ModMax BH.  In section \ref{sect:NullGeodesics} we compare the light orbits of the two effective metrics of the ModMax BH with the massless trajectories in the Reissner Nordstrom (RN) BH. The deflection angles and the gravitational lensing are calculated in section \ref{sect:Lensing}. The redshifts produced by the two effective metrics are determined, first considering an emitter moving in a circular orbit and then a static emitter. In section \ref{sect:shadow} the shadow of the ModMax BH is determined and we set bounds on the values of the nonlinear parameter $\gamma$ that are consistent with the observations of the shadow of the BH at the center of our galaxy, Sagitarius $A^{\ast}$.
We compare the aforementioned effects in the ModMax BH vicinity with the corresponding ones of the RN BH, its linear counterpart. 

We use $G=c=1$ and the signature of the metric is $(-,+,+,+)$. 
\section{Effective or optical metric for light trajectories in NLED.}
\label{sect:EffectiveMetric}

NLED theories are characterized by a Lagrangian $L(F,G)$ that depends nonlinearly on the two electromagnetic Lorentz invariants $F=F_{\mu\nu}F^{\mu\nu}$ and $G=F_{\mu\nu}^{*}F^{\mu\nu}$, with $F^{*\mu\nu}=\frac{1}{2}\epsilon^{\mu\lambda\alpha\beta}F_{\alpha \beta}$ being the dual of $F_{\mu\nu}$. In an environment with very intense electromagnetic fields light trajectories are not the null geodesics of the geometric or background metric but are the null geodesics of an effective or optical metric.  The effective metric is determined from the analysis of the propagation of the electromagnetic field discontinuities through characteristic surfaces \cite{Pleban, Novello2000b}. It has been shown that the effective metric approach is equivalent to the soft photon approximation \cite{Novello2000}. Considering a null vector $k_{\mu}=(\omega,\vec{k})$  that is normal to the characteristic surfaces or wavefronts; the effective  metric $g_{\rm eff}^{\mu \nu}$ must be the metric in which the wave vector is null, $k^{\mu}k_{\mu}=0,$
 \begin{equation}
     g_{{\rm eff} (a)}^{\mu\nu}k_{\mu}k_{\nu}=0, \quad a=1,2
 \label{nullgeod1}
 \end{equation}
where the subscript $a=1,2$  corresponds to the two possible effective metrics that can arise in NLED, and that give place to the phenomenon of birefringence, i.e. there are two possible light trajectories, depending on the light polarization.
As was mentioned before, 
NLED can be modeled as a material medium characterized by indices of birefringence or refractive indices \cite{Novello2000b, Bandos2020}, $\lambda_{a}$,  given in terms of the NLED Lagrangian and its derivatives as,

\begin{equation}
    \lambda_{a}=-4\frac{L_{FF}+\Omega_{(a)}L_{FG}}{L_{F}+G(L_{FG}+\Omega_{(a)} L_{GG})},
    \label{refindices}
\end{equation}
where  ${L}_{X}=\frac{d\mathcal{L}}{dX}$, $X={F, G}$, and $\Omega_{(a)}$ depends on the derivatives of the Lagrangian with respect to the invariants \cite{Novello2000b}.  The expression for the effective metric can be identified from the dispersion relation, Eq. (\ref{nullgeod1}), 

\begin{equation}
   \left\{ g^{\mu\nu} + \lambda_{a}t^{\mu\nu} \right\} k_{\mu}k_{\nu}= 
   g_{{\rm eff}  (a)}^{\mu\nu}k_{\mu}k_{\nu}=0,
\label{nullgeod2}
\end{equation}
where $g^{\mu\nu}$ is the background metric and $t^{\mu \nu}=F^{\mu \lambda }F_{\lambda}{}^{\nu}$. 
And  the effective metrics for general nonlinear electrodynamics with Lagrangian $L(F,G)$ are given by,
\begin{equation}
  g_{{\rm eff} (a)}^{\mu\nu} =  g^{\mu\nu}-4\frac{L_{FF}+\Omega_{(a)}L_{FG}}{L_{F}+G(L_{FG}+\Omega_{(a)} L_{GG})}t^{\mu\nu}.
\label{eff_metric}
\end{equation}
\subsection{Phase velocity}

The phase velocity of light can be calculated
from the dispersion relation.  Let us consider a  null vector $k_{\mu}=\{ \omega, \vec{k} \}$. If $k_{i}$ is the light propagation  direction, from the dispersion relation,  Eq. (\ref{nullgeod1}), 

\begin{equation}
g_{{\rm eff}(a)}^{tt}\omega^{2}+2g_{{\rm eff}(a)}^{it}\omega k_{i}+g_{{\rm eff}(a)}^{ij} k_i k_j=0,
\label{dr1}    
\end{equation}
where the latin subscripts (or superscripts) denote the spatial coordinates. Defining the normalized wave vector $\tilde{k}_{i}=k_{i}/|\vec{k}|$, then Eq. (\ref{dr1}) can be written  as 

\begin{equation}
    g_{{\rm eff}(a)}^{tt}\frac{\omega^{2}}{|\vec{k}|^2}+2g_{{\rm eff}(a)}^{it}\frac{\omega}{|\vec{k}|} \tilde{k}_{i}+g_{{\rm eff}(a)}^{ij} \tilde{k}_{i}\tilde{k}_{j} =0.
\end{equation}

Then the light's phase velocity,  $v=\omega/|\vec{k}|$, for propagation in the direction $\tilde{k}_{i}$ $v^{i}$ is given by 
\begin{equation}
   \left(v^{i} \right)_{a}= \frac{\omega}{|\vec{k}|}\tilde{k}_{i}=\frac{g_{{\rm eff}(a)}^{it}\tilde{k}_{i}}{g_{{\rm eff}(a)}^{tt}}\pm\sqrt{\left(\frac{g_{{\rm eff}(a)}^{it}\tilde{k}_{i}}{g_{{\rm eff}(a)}^{tt}}\right)^2-\frac{g_{{\rm eff}(a)}^{ij}\tilde{k}_{i}\tilde{k}_{j}}{g_{{\rm eff}(a)}^{tt}}}, \qquad a=1,2
    \label{phase_velocity}
\end{equation}

in the case that the effective metric is diagonal, i.e. $g_{ti}=0$ and $g_{ij}=0, \quad i \ne j$,  Eq. (\ref{phase_velocity}) simplifies to $\left(v^{i} \right)_{a}=\pm\sqrt{-\frac{g_{{\rm eff}(a)}^{ij}\tilde{k}_{i}\tilde{k}_{j}}{g_{{\rm eff}(a)}^{tt}}}$.

The two possible effective metrics $g_{{\rm eff}(a)}^{\mu\nu}$, $a=1,2$ render two
dispersion relations that correspond to two modes of polarization, this is known as the birefringence effect \cite{Novello2000}. Note that these velocities differ from the phase velocities of a massless test particle that follows the null geodesics of the background metric $g_{\mu\nu}$; they are given by analogous expressions making
$g_{\rm eff(a)}^{\mu\nu} \mapsto g^{\mu\nu}$.

\section{ModMax Nonlinear electrodynamics.}
\label{sect:ModMax}
The ModMax nonlinear electrodynamics possesses both Maxwell's symmetries, conformal invariance and SO(2) duality-rotation invariance.
The  ModMax NLED Lagrangian was derived in  \cite{Kosyakov2020} using the Bessel-Hagen criterion for conformal invariance, and the Gaillard–Zumino one for invariance under duality transformations.
In \cite{Bandos2020} the symmetries of the theory are deduced by means of the Hamiltonian formalism and it is demonstrated that there are only two theories that share the same symmetries as Maxwell's, the Bialynicki-Birula (BB) and the ModMax. Considering that the Hamiltonian depends on two parameters, one with dimensions of energy density ($T$) and the other being the dimensionless parameter $\gamma$, the BB theory is the generalization for strong fields ($T\rightarrow0$) and cannot be written in Lagrangian form \cite{Bandos2020}. The ModMax theory is the generalization for weak fields ($T \rightarrow \infty$) and Maxwell's theory is recovered when the nonlinear parameter $\gamma$ vanishes. 

The ModMax Lagrangian is given by

\begin{equation}
    \mathcal{L}_{\rm ModMax}=-\frac{F}{4}\cosh{\gamma}+\frac{\sinh{\gamma}}{4}\sqrt{F^2 +G^2},
    \label{lagrangianModMax}
\end{equation}
where $F$ and $G$ are the electromagnetic Lorentz invariants; the Maxwell Lagrangian \cite{jackson}, $\mathcal{L}_{\rm M}=-\frac{F}{4}$,  corresponds to the vanishing of the nonlinear parameter $\gamma=0$ (specifics of the ModMax theory can be found in \cite{Kosyakov2020, Pergola2021}). The birefringence indices $\lambda_{a}$ in Eq. (\ref{refindices}), for the ModMax Lagrangian, Eq. (\ref{lagrangianModMax}), are

\begin{equation}
\lambda_{1}=\frac{L_{FF}+L_{GG}}{L_{F}+2\left(G L_{FG}-F L_{GG} \right)}, \qquad \lambda_{2}=0, 
\end{equation}
or in terms of $\gamma$ and the electromagnetic invariants $F$ and $G$,
\begin{equation}
\lambda_{1}=\frac{4\tanh{\gamma}}{\sqrt{F^2 + G^2}+F \tanh{\gamma}},
      \qquad 
     \lambda_{2}=0.
     \label{refind12}
\end{equation}

The effective metric $g_{\rm eff  (1)}^{\mu\nu} $ from Eq. (\ref{eff_metric}) is given by
\begin{equation}
\label{eff_metric_1}
g_{{\rm eff}(1)}^{\mu\nu}= g^{\mu\nu} +\frac{4\tanh{\gamma}}{\sqrt{F^2 + G^2}+F \tanh{\gamma}}t^{\mu\nu}.
\end{equation}

As a consequence of the conformal invariance  one of the  birefringence index vanishes, $\lambda_{2}=0$, then the corresponding effective metric is equal to the background metric,

\begin{equation}
g_{{\rm eff}(2)}^{\mu\nu} = g^{\mu\nu},
\label{eff_metric_2}
\end{equation}
this means that one of the polarization modes follows the null geodesics of the background metric, with the dispersion relation given by $\omega^2=|\vec{k}|^2$.

The existence of two effective metrics corresponds to the birefringence effect, i.e. there are two possible paths that light rays can follow, depending on their polarization. In case $\gamma=0$ we recover one single effective metric for the propagation of electromagnetic waves $g^{\mu\nu}_{{\rm eff}(1)}=g^{\mu\nu} = g^{\mu\nu}_{{\rm eff}(2)}$.

Let us, for instance, consider a vanishing electric field and a uniform magnetic field $\vec{B}$ in Minkowski space as the background spacetime, $g^{\mu\nu}=\eta^{\mu\nu}$, then the dispersion relation amounts to  
\begin{equation}
    \omega^2 = k^2 (\cos^2\phi + {\rm e}^{-2\gamma} \sin^2\phi)
\end{equation}
where $\phi$ is the angle between the propagation direction $\vec{k}$ and the magnetic field, $\vec{B}$. There is no birefringence when $\phi=0$ because in this case, the background preserves the rotational symmetry in the plane defined by $\vec{k}$ \cite{Bandos2020}. This equation also indicates that superluminal velocities are reached for negative values of the nonlinear parameter $\gamma$; to avoid superluminal velocities we should restrict to $\gamma \ge 0$.

In case the effective metric  is  diagonal, the two phase velocities, from Eq. (\ref{phase_velocity}), are

\begin{eqnarray}
\label{phasevel1}
(v_{1})^2=\left(\frac{\omega}{|\vec{k}|} \right)^2_{1} &=& -\frac{ g^{ij}+ \lambda_1 t^{ij}}{g^{tt}+ \lambda_1 t^{tt}}\tilde{k}_{i}\tilde{k}_{j},\\
(v_{2})^2=\left(\frac{\omega}{|\vec{k}|} \right)^2_{2} &=& -\frac{g^{ij}}{g^{tt}}\tilde{k}_{i}\tilde{k}_{j}.
\label{phasevel2}
\end{eqnarray}

In the next section, we present the ModMax NLED BH and analyze several aspects of light propagation.

\section{ModMax black hole}
\label{sect:ModMaxBH}

In \cite{Maceda2020}  was derived the SSS solution to the Einstein equations coupled to ModMax NLED; it is characterized by the BH mass, BH charge, and the nonlinear parameter $\gamma$; its line element is given by 
\begin{equation}
    ds^2 = g_{\mu\nu} dx^{\mu}dx^{\nu}= -f(r)dt^2+\frac{1}{f(r)}dr^2+r^2d\Omega, \quad
    f(r)=1-\frac{2 M}{r}+\frac{e^{-\gamma } Q^2}{r^2},
    \label{bhmetric}
\end{equation}
 where $d\Omega=d\theta^2+ \sin^2{\theta} d\phi^2$; the charge can be electric $Q= Q_{e}$,  magnetic  $Q=Q_{m}$, or both (Dyonic case)  $Q=\sqrt{Q_{e}^2 + Q_{m}^2}$, \cite{Maceda2020}. This metric  represents a charged BH with horizons defined by the roots of $f(r)=0$, 
\begin{equation}
    r_{+}=M+\sqrt{M^2- e^{-\gamma }Q^2}, \quad  r_{-}=M-\sqrt{M^2-e^{-\gamma }Q^2}.
    \label{eventhorizons}
\end{equation}
There exists an event horizon $r_{+}$ if the condition  $0 \leq Q^2 e^{-\gamma } \leq M^2$ is fulfilled. 
We shall consider $g_{\mu\nu}$ in Eq. (\ref{bhmetric}) as the background metric. Recalling the metric function for the RN BH,
  \begin{equation}
 f_{RN}(r)=1-\frac{2 M}{r}+\frac{Q^2}{r^2},
 \end{equation}
 note that the ModMax metric function $f(r)$ resembles the RN one with a screened charge.
 In the RN BH the charge is restricted to $Q^2 \le M^2$, while the ModMax BH can bear a larger charge, $Q^2 \leq e^{\gamma } M^2$, due to the charge screening.
 
\begin{figure}[H]
    \centering
    \includegraphics{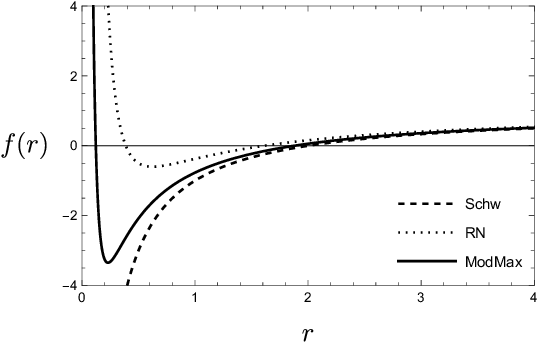}
    \caption{The metric functions $f(r)$ of Schwarszchild, RN and  ModMax BHs are shown. The radii of the event horizons are the intersections of the corresponding metric functions with the $r-$axis: the radius of the event horizon for the ModMax BH is larger than the RN one and smaller than Schwarszchild's. In this plot, we fixed charge $Q=0.8$ and  $\gamma=0.5$.}
    \label{fig:Fig1}
\end{figure}

Since the nonvanishing components of the electromagnetic tensor are $F_{rt}=Q_{e}/r^2$ and $F_{\phi \theta}= Q_{m}\sin{\theta}$, the tensor $t^{\mu\nu}$ in Eq. (\ref{eff_metric_1}) is given by

\begin{equation}
t^{\mu\nu}=F^{\mu\lambda}F_{\lambda}{}^{\nu}= {\rm diag} \left[ -\frac{F_{tr}^2}{f(r)}, f(r)F_{tr}^{2},- \frac{F_{\phi \theta}^2}{r^6},-\frac{F_{\phi \theta}^2}{r^6} \right].
\label{tmunu}
\end{equation}
In Fig. \ref{fig:Fig1} we compare the metric functions $f(r)$ of the Schwarszchild, the RN, and the ModMax BHs. The radius of the ModMax BH event horizon is larger than the RN one and smaller than Schwarszchild's.  
The ModMax BH is singular at $r=0$, i.e. its curvature scalars diverge at $r=0$.
In the next subsections, we analyze light trajectories in the vicinity of the ModMax BH, by calculating the null geodesics of the effective metrics.
\subsection{Phase velocities in the vicinity of ModMax BH}
\label{sect:Birefringence}

Our aim in this subsection is to determine the
phase velocities of light propagating near the ModMax BH.
Without loss of generality, we consider equatorial trajectories of light, $\theta= \pi/2$. Then considering the tensor $t^{\mu \nu}$ in Eq. (\ref{tmunu}), the effective metrics, Eqs. (\ref{eff_metric_1}) and (\ref{eff_metric_2}),   are given by

\begin{equation}
g_{{\rm eff}(1) }^{\mu\nu}=\frac{Q^2}{e^{-\gamma} Q_{e}^2+e^{\gamma}Q_{m}^2}  {\rm diag} 
    \left[
 -e^{\gamma }f(r)^{-1} , e^{\gamma } f(r) ,  \frac{e^{-\gamma }}{r^2},  \frac{e^{-\gamma }}{r^2} 
\right],
\label{effmetric1}
\end{equation}

\begin{equation}
g_{{\rm eff}(2)}^{\mu \nu}= g^{\mu \nu}={\rm diag}\left[
 -f(r)^{-1} , f(r), \frac{1}{r^2}, \frac{1}{r^2} \right],
\label{effmetric2}
\end{equation}
where $f(r)=1-\frac{2 M}{r}+\frac{e^{-\gamma } Q^2}{r^2}$, $Q^2=Q_{e}^2+Q_{m}^2$; and we have used  the birefringence indices, that for the ModMax BH, are,

\begin{equation}
\lambda_1= \frac{2 e^{\gamma} r^4 \sinh{\gamma}}{e^{2 \gamma} F_{\phi \theta}^2+r^4 F_{r t}^2}=
 \frac{2  r^4 \sinh{\gamma}}{e^{- \gamma} Q_{e}^2+ e^{\gamma} Q_{m}^2}, \quad \lambda_2=0.
\end{equation}

Note that both metrics are diagonal, then the expression of the phase velocity simplifies. 
Considering the propagation along radial and $\phi$-angular directions, with a wave vector given by $k_{\mu}=(\omega,1,0,1),$ using Eq. (\ref{phase_velocity})

\begin{equation}
   \left(v^{r\phi} \right)_{a}=\pm\sqrt{-\frac{g_{{\rm eff}(a)}^{rr}\tilde{k}_{r}^2+g_{{\rm eff}(a)}^{\phi\phi}\tilde{k}_{\phi}^2}{g_{{\rm eff}(a)}^{tt}}}, \qquad a=1,2
\end{equation}

Then, for the two values of $\lambda_{1,2}$ in Eq. (\ref{refind12}) the corresponding phase velocities are
\begin{equation}
    v^{r\phi}_{1}=\sqrt{f(r) \left(f(r)+\frac{e^{-2 \gamma }}{r^2}\right)},
    \label{phasevelr21}
\end{equation}

\begin{equation}
    v^{r\phi}_{2}=\sqrt{f(r) \left(f(r)+\frac{1}{r^2}\right)}.
    \label{phasevelr22}
\end{equation}

For light propagating in a purely radial direction,  ($\phi$ and $\theta$ fixed), the two phase velocities turn out to be equal, then there is no birefringence. From Eq. (\ref{phasevel1}),

\begin{equation}
    v^{r}_{1,2}=f(r).
    \label{radialphasevel0}
\end{equation}

The light phase velocity in the neighborhood of the ModMax BH is always less than the one corresponding to RN BH, $v^{r}_{RN} > v^{r}_{1,2}$. 
If $v^r=0$ the orbits are circular, and if $\theta=\pi/2$, the light ray stays in the equatorial plane, however, these are unstable circular orbits (UCO).
If the wave vector has angular components $k^{\theta}$ or $k^{\phi}$, then there is birefringence, $v^{i}_{1} \ne v^{i}_{2}$,

$\theta$-direction, $i=\theta$:

\begin{equation}
v_{1}^{\theta}=\frac{{\rm e}^{-\gamma}}{r}\sqrt{f(r)}, \qquad
v_{2}^{\theta}=\frac{1}{r}\sqrt{f(r)},
\label{phaseveltheta}
\end{equation}

$\phi$-direction, $i=\phi$:

\begin{equation}
v_{1}^{\phi}=\frac{{\rm e}^{-\gamma}}{r\sin{\theta}}\sqrt{f(r)}, \qquad
v_{2}^{\phi}=\frac{1}{r \sin{\theta}}\sqrt{f(r)}.
\label{phasevelphi}
\end{equation}

\begin{figure}[H]
    \centering
    \includegraphics[width=0.6\textwidth]{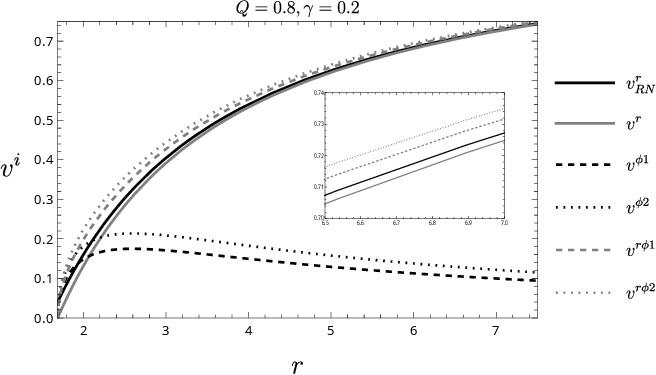}
    \caption{The radial and $\phi$-angular phase velocities corresponding to the two effective metrics are shown.  Measured by a distant observer, starting from infinity with velocity $c=1$, as light approaches the horizon its phase velocity tends to zero.	There is no birefringence in the radial direction; if the wave vector has angular components then there is birefringence. The gray curve is the phase velocity in the radial direction. The black dotted and dashed curves correspond to the phase velocity in purely $\phi$-direction $v^{\phi2}$, $v^{\phi1}$ with $v^r=0$. The gray dashed curve corresponds to the phase velocity $v^{r \phi}$ considering that the light propagates along radial and $\phi$-angular directions.
The phase velocities in the neighborhood of the RN BH (black solid curve for the radial direction, angular directions are not shown) are always greater than the ones for the ModMax BH. We fixed the charge and nonlinear parameter as $Q=0.5$, $\gamma=0.2$. }
    \label{fig:Fig2}
\end{figure}

In Figure \ref{fig:Fig2} we compare the phase velocities in equatorial orbits. 
The phase velocities depend on the BH charge $Q$ through the metric function $f(r)$ and the occurrence of birefringence is independent of whether the BH is electrically or magnetically charged.
The effect of the nonlinear parameter $\gamma$ is to reduce the phase velocities.

\subsection{Light trajectories, null Geodesics, and orbits around the ModMax BH. }
\label{sect:NullGeodesics}
We consider the Hamiltonian formalism and the conserved quantities for a test particle to obtain the equations of motion of a photon in the external field of a BH.
In a static spherically symmetric metric (SSS), due to the existence of two Killing vectors, a test particle has two conserved quantities, its energy, and its angular momentum.
\begin{equation}
    E=g_{tt}\frac{dt}{d\tau}, \qquad L=g_{\phi\phi}\frac{d\phi}{d\tau}.
    \label{conservedq}
\end{equation}
where $\tau$ is the affine parameter along a geodesic. Without loss of generality, we consider equatorial orbits, $\theta=\pi/2$. Then from the mass invariance,
$g_{\mu \nu} \dot{x^{\mu}} \dot{x^{\nu}}= \delta$, and making a distinction between the background metric and the effective metric,

\begin{equation}
   \left(\frac{dr}{d\tau} \right)^2 
   + \frac{1}{g^{{\rm eff}(a)}_{rr}} \left( g^{{\rm eff}(a)}_{\phi\phi}\left( \frac{L}{g_{\phi\phi}} \right)^2
   +g^{{\rm eff}(a)}_{tt}\left( \frac{E}{g_{tt}}\right)^2 
   - \delta \right)
   = 0,
\end{equation}
where $\delta= 1,0,-1$ for space-like, null, or time-like geodesics, respectively; 
denoting $\dot{r}=dr/d\tau$ and  the impact parameter $b= L/E$, the previous equation is
\begin{equation}
   \dot{r}^2 
   + \frac{L^2}{g^{{\rm eff}(a)}_{rr}} \left( \frac{g^{{\rm eff}(a)}_{\phi\phi}}{r^4}
   +\frac{g^{{\rm eff}(a)}_{tt}}{f^2 (r)} \frac{1}{b^2} 
   - \frac{\delta}{L^2} \right)
   = 0,
   \label{geodesic}
\end{equation}
and $a=1,2$ denote the two effective metrics. Considering 
$\dot{r}^2 + V^{(a)}_{\rm eff} =0,$ we identify the effective potential as 

\begin{equation}
    V^{(a)}_{\rm eff}=\frac{L^2}{g^{{\rm eff}(a)}_{rr}} \left( \frac{g^{{\rm eff}(a)}_{\phi\phi}}{r^4}
   +\frac{g^{{\rm eff}(a)}_{tt}}{f^2 (r)} \frac{1}{b_{a}^2} 
   - \delta \right).
   \label{effpotential}
\end{equation}

The radius of the  circular orbits $r_{c}$ can be determined  from the conditions on the effective potential,  
\begin{equation}
V^{(a)}_{\rm eff}=0, \quad  \frac{dV^{(a)}_{\rm eff}}{dr}=0,
\label{circ_orb_conds}
\end{equation}
and considering that the photons propagate along null geodesics of the effective metric ($\delta=0$), we 
obtain  the  radius of the circular orbits $r_c$ as
 
\begin{equation}
    r_{c}=\frac{3M}{2}\left(1 \pm \sqrt{1 - \frac{8 e^{-\gamma } Q^2}{9 M^2}} \right).
    \label{criticalradius}
\end{equation}
There are two impact parameters, $b_{c}^{a}$,  corresponding to the two effective metrics, given by

\begin{equation}
 \left( b^{2}_{c} \right)_{a} = \left( -\frac{r_{c}^4 g^{{\rm eff}(a)}_{tt}}{f^{2}(r_{c})g^{{\rm eff}(a)}_{\phi\phi}}\right)_{r_{c}}, \qquad a=1,2
    \label{critimpactparam}
\end{equation}
or explicitly, 
\begin{equation}
    b_{c1}^2=\frac{e^{-2\gamma}r_{c}^2}{f(r_{c})}, \qquad 
    b_{c2}^2=\frac{r_{c}^2}{f(r_{c})}.
    \label{impactparam12}
\end{equation}
From Eq. (\ref{geodesic}) and  $\dot{\phi}= L/g_{\phi \phi}$ we can write the equation for the light  trajectories in the $r-\phi$ plane, $\phi_{a}(r)$,  

\begin{equation}
   \left(\frac{dr}{d\phi}\right)_{a}^{2} = \left( \frac{\dot{r}}{\dot{\phi}} \right)_{a}^2 =-  \frac{1}{g^{{\rm eff}(a)}_{rr}} \left( g^{{\rm eff}(a)}_{\phi\phi}+g^{{\rm eff}(a)}_{tt}\frac{r^4}{f^2(r)}\frac{1}{b_{c}^2} \right)
\end{equation}
and using the value of the critical impact parameters $b_{c a}$, Eq. (\ref{critimpactparam})
\begin{equation}
    \left(\frac{dr}{d\phi}\right)_{a}^2 =-  \frac{g^{{\rm eff}(a)}_{\phi\phi}}{g^{{\rm eff}(a)}_{rr}}\left(1
    -\frac{g^{{\rm eff}(a)}_{tt}}{g^{{\rm eff}(a)}_{\phi\phi}}\left(\frac{g_{\phi\phi}^{{\rm eff}(a)}}{g^{{\rm eff}(a)}_{tt}} \right)_{r_{c}} \frac{r^4}{r_{c}^4} 
 \frac{f^2(r_{c})}{f^2(r)}\right),
 \label{integralrphi}
\end{equation}
where $f(r)$ is given in Eq. (\ref{bhmetric}). Following a standard procedure to integrate the previous equation \cite{Chandrasekhar1985}, the $r$ coordinate is transformed as $r=1/u$. 
In our case there are two possible trajectories of the photon $\phi_{a}(r)$, $a=1,2$; for $\phi_{1}(r)$ we have

\begin{equation}
    \left(\frac{d u}{d \phi_{1} }\right)^2=e^{2 \gamma } (u-u_{c})^2 \left[ 2 u \left(M-e^{-\gamma}Q^2 u_{c}\right)+u_{c} \left(M-e^{-\gamma}Q^2 u_{c}\right)-e^{-\gamma}Q^2 u^2\right],
    \label{orbiteq}
\end{equation}
where $u_{c}=1/r_{c}$. Then it has to be integrated 
\begin{equation}
    {\rm e}^{\gamma}d\phi_{1}=\pm \int \frac{d\xi}{\sqrt{-e^{-\gamma}Q^2 + c_1 \xi + c_2  \xi^2}}
\end{equation}
where 
\begin{equation}
    \xi=\frac{1}{u-u_{c}}, \quad c_{1}=2(M-2e^{-\gamma}Q^2 u_{c}), \quad c_{2}=u_{c}\left(3M-e^{-\gamma}Q^2 u_{c}\right),
\end{equation}
and
\begin{equation}
    u_{c}= \frac{3M}{4e^{-\gamma}Q^2}\left(1-\sqrt{1-\frac{8e^{-\gamma}Q^2}{9M^2}} \right).
\end{equation}

The solutions for $\phi_{1}(r)$ depend on the sign of $c_{2}$,
\begin{eqnarray}
\label{phic2p}
 (\phi_{1})_{\pm} & = & \mp {\rm e}^{-\gamma} \left(\frac{1}{\sqrt{c_{2}}} \ln{\left(c_{1}+2c_{2}\xi+2\sqrt{c_{2}}\sqrt{-e^{-\gamma}Q^2+c_{1}\xi +x \xi^2 }\right)} \right), \quad c_{2}>0\\
(\phi_{1})_{\pm} & = & \mp {\rm e}^{-\gamma} \left(-\frac{1}{\sqrt{c_{2}}}\arcsin{\left(\frac{2 c_{2} \xi + c_{1} }{\sqrt{4 e^{-\gamma}Q^2 c_{2} + c_{1}^2}} \right)} \right), \quad c_{2}<0
\end{eqnarray}
where the ${\pm}$ sign corresponds to the angle measured counterclockwise or clockwise; we consider the positive solution. 

We can clear out $r(\phi)$  from Eq. (\ref{phic2p}),

\begin{equation}
    r_{1}(\phi_1)= \frac{\left(\left(c_{1}-{\rm Exp}(\pm {\rm e}^{\gamma}\sqrt{c_{2}}\phi)\right)^2 + 4 c_{2} e^{-\gamma}Q^2 \right)r_{c}}{\left(\left(c_{1}-{\rm Exp}(\pm {\rm e}^{\gamma}\sqrt{c_{2}}\phi)\right)^2 + 4 c_{2} e^{-\gamma}Q^2 \right) +  4 c_{2} r_{c}{\rm Exp}(\pm {\rm e}^{\gamma}\sqrt{c_{2}}\phi)}, \qquad c_{2}>0
\end{equation}

Following an analogous procedure, we determine  the second photon trajectory, $\phi_{2}(r)$, corresponding to the second effective metric, as
\begin{eqnarray}
\phi_{2}(r) & = & \mp \left(\frac{1}{\sqrt{c}} \ln{\left(c_{1}+2c_{2}\xi+2\sqrt{c_{2}}\sqrt{-e^{-\gamma}Q^2+c_{1} \xi + c_{2}\xi^2 }\right)} \right), \quad c_{2}>0\\
\phi_{2}(r) & = & \mp \left(-\frac{1}{\sqrt{c_{2}}}\arcsin{\left(\frac{2 c_{2} \xi + c_{1} }{\sqrt{4 e^{-\gamma}Q^2 c_{2} + c_{1}^2}} \right)} \right), \quad c_{2}<0
\end{eqnarray}

and the trajectory   $r_{2}$ as a function of $\phi_2$ is 

\begin{equation}
    r_{2} (\phi_2)= \frac{\left(\left(c_{1}-{\rm Exp}(\pm \sqrt{c_{2}}\phi)\right)^2 + 4 c_{2} e^{-\gamma}Q^2 \right)r_{c}}{\left(c_{1}-{\rm Exp}(\pm \sqrt{c_{2}}\phi)\right)^2 + 4 c_{2} \left( e^{-\gamma}Q^2 + r_{c}{\rm Exp}(\pm \sin{\theta}\sqrt{c_{2}}\phi)\right)}.
\end{equation}

\begin{figure}[H]
    \centering
    \includegraphics[width=0.65\textwidth]{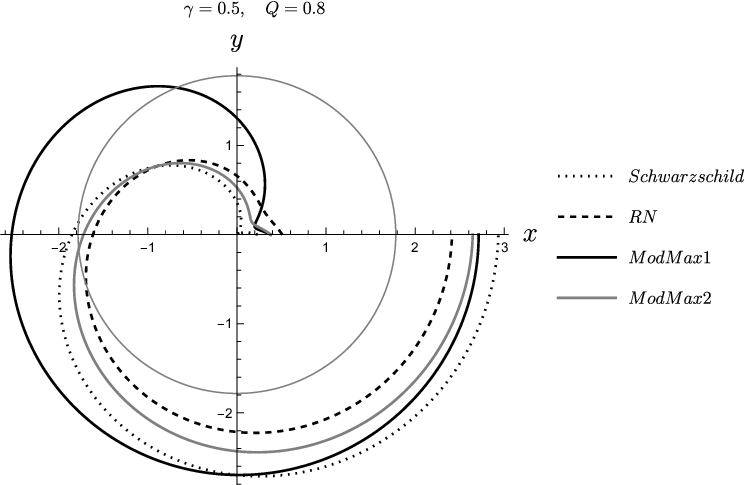}
    \caption{Photon trajectories in the vicinity of the Schwarzschild (dotted), RN (dashed), and ModMax  (solid)  BHs are shown. The gray circumference indicates the ModMax event horizon;  the trajectories inside the event horizon are not considered in our study. The trajectory of the effective metric that coincides with the background ModMax metric lies between the Schwarzschild and the RN ones. In this plot are fixed  $M=1$, $Q=0.8$, $\gamma=0.5$. }
     \label{fig:Fig3}
\end{figure}

The expression for the latter trajectory, $r_2(\phi_2)$, is identical to the one in the vicinity of an RN BH, but for a reduced charge, $Q \mapsto Q \exp{- \gamma /2}$; this orbit is as well the one corresponding to the massless test particles.
In Fig. \ref{fig:Fig3} we show radial orbits for the ModMax, RN, and Schwarzschild BHs. The trajectory corresponding to the metric  $g^{{\rm eff}(2)}_{\mu\nu}$, that coincides with the background metric lies between the Schwarzschild and RN trajectories, this is because the only difference with RN BH  is the charge screening. In contrast, the null geodesics of  $g^{{\rm eff}(1)}_{\mu\nu}$, have larger radius, then the traveled distance before crossing the horizon is also larger.  
As $\gamma$ increases the screening of the charge increases and the orbit approaches the Schwarzschild one, this is, the unstable circular orbit of radius $3M$ characteristic of the Schwarzschild BH photosphere.
\subsection{Lensing and deflection angle}
\label{sect:Lensing}

Gravitational lensing is a relativistic phenomenon consisting of the deflection of light rays in the vicinity of a massive object; it can be used to survey massive dark objects both in weak and strong gravitational fields. Many aspects of this effect produced by BH have been reported in the literature:  the exact lens equation for the Einstein-Euler-Heisenberg static black hole \cite{Amaro2022},  the RN BH and RN-de Sitter BH lensing  \cite{Eiroa2002, Zhao2016}, the lensing in the strong field limit  \cite{Bozza2002, Tsukamoto2021, Jia2021} and the gravitational lensing of massive particles in RN BH \cite{Pang2019}.  Recently the study of the lensing and the shadow of the ModMax BH was presented in \cite{Pantig2022}, where the massless particle trajectories are determined from the background metric; such that their results correspond to ours for the effective metric that coincides with the BH metric, $g^{{\rm eff}(2)}_{\mu\nu}= g_{\mu\nu}$.  Therefore the deflection of light calculated from the effective metric $g^{{\rm eff}(1)}_{\mu\nu}$  complement the ones in \cite{Pantig2022}. Moreover, in the weak field limit,  we present a more precise expression for the deflection angle including terms of higher order in $Q$. The results in  \cite{Pantig2022}  for the weak deflection angle and the Einstein ring of the ModMax BH are compared with the data obtained from Sagittarius $A^{*}$(Sgr $A^{*}$) and M87*, then they can be a reference for ours as well.

\begin{figure}[H]
    \centering
    \includegraphics[width=0.6\textwidth]{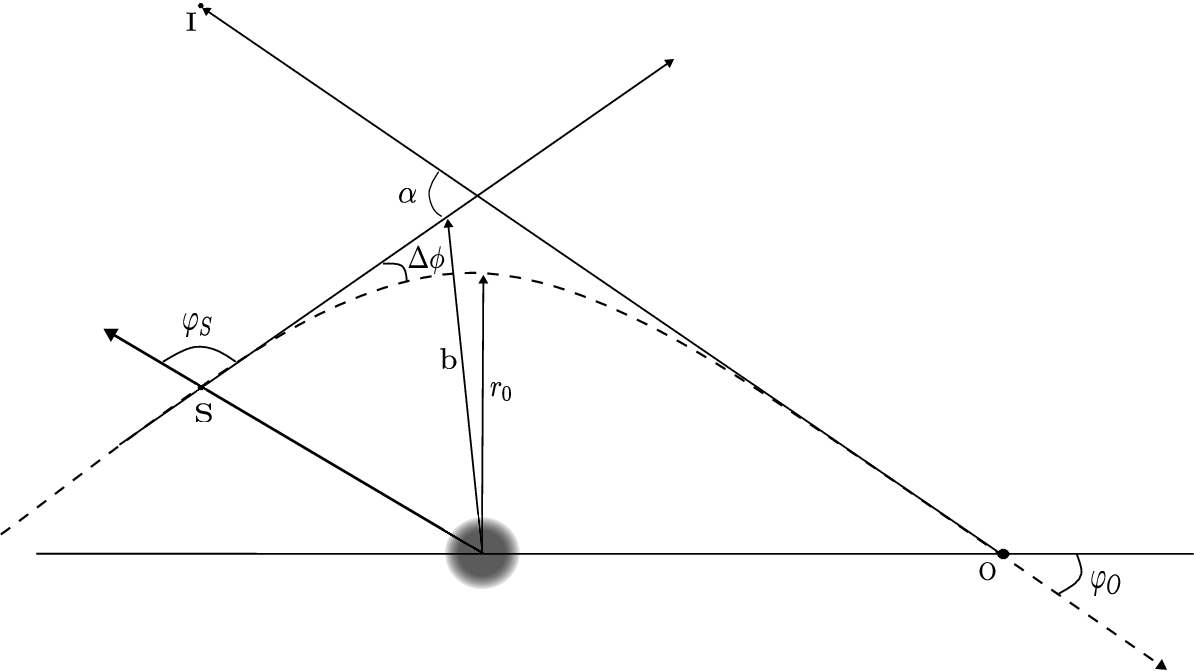}
    \caption{The diagram shows  the light deflection angle  $\alpha$,  the position of the source and of the observer,  $S$ and $O$, respectively. $b$ is the impact parameter and $r_{0}$ the distance of closest approach. The gray circle indicates the position of the BH acting as a lens. }
    \label{fig:Fig4}
\end{figure}

The expression for the deflection angle $\alpha$ produced by the presence of a BH acting like a lens in the light trajectory, given  in terms of the  distance of closest approach, denoted by  $r_{0}$, is

\begin{equation}
    \alpha(r_{0}) = I(r_{0})+\varphi_{O}-\varphi_{S}
\end{equation}

where the angles $\varphi_{O}-\varphi_{S}$ are illustrated in Fig. \ref{fig:Fig4} and $I(r_{0})$ is calculated from the orbit Eq. (\ref{integralrphi}) considering that the observer and the source are in the same plane of a flat space-time region and that the source of light is located at infinity,  $I(r_0)=2\Delta \phi=2 \left[ \phi (r \mapsto {\infty})-\phi({r_{0}}) \right]$, \cite{Weinberg1972}, and  $I(r_{0})$ is given by 

\begin{equation}
    I_{a}(r_{0})=2\int_{r_{0}}^{\infty}
    \left(\frac{d\phi_{a}}{dr} \right)dr, \quad a= 1,2.
 \label{int1}
\end{equation}
where $\left(\frac{d\phi}{dr} \right)$ can be derived from Eq. (\ref{integralrphi}).
If the trajectory were a straight line as in Minkowski spacetime, there would be no deflection and $\varphi_{O}-\varphi_{S}=-\pi$, but the presence of the BH bends light's trajectory and 
$\varphi_{O}-\varphi_{S} \ne -\pi$
and to determine the difference we need to calculate the limit $r_{0}\rightarrow \infty$ in equation (\ref{int1}), making  $I_{r_{0}}=0$.  
For the effective metrics in consideration, we obtain two deflection angles, $\alpha_{1}$, $\alpha_{2}$, that are given by

\begin{equation}
  \alpha_{1}(r_{0})=   I_{1}(r_{0})=  e^{-\gamma } \left\{ \int_{r_{0}}^{\infty} \frac{2}{r\sqrt{\frac{r^2}{r_{0}^2} f(r_{0})-f(r)}}dr - \pi \right\}= e^{-\gamma } \alpha_2.
\end{equation}

To integrate $I_{r_{0}}=0$   we transform  to the variable $z$,  $z=1-\frac{r_{0}}{r}$ obtaining
\begin{equation}
    I_{1}(r_{0})= e^{-\gamma}  \int_{0}^{1}{2 \left( f(r_{0})-\left(1-z \right)^2 \left(1-\frac{2M}{r_{0}}(1-z)+\frac{Q^2 (1-z)^2}{r_{0}^2} \right) \right)^{-1/2}dz}.
\end{equation}

\begin{figure}[H]
    \centering
    \subfigure{\includegraphics[width=0.4\textwidth]{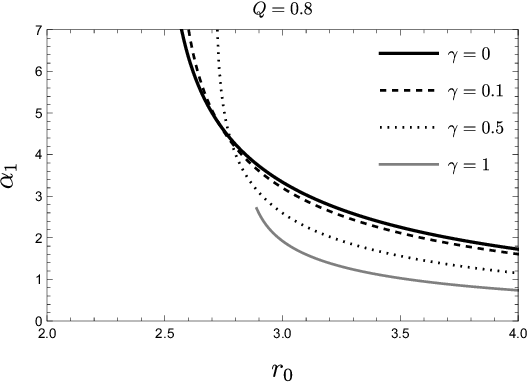}}
    \subfigure{\includegraphics[width=0.4\textwidth]{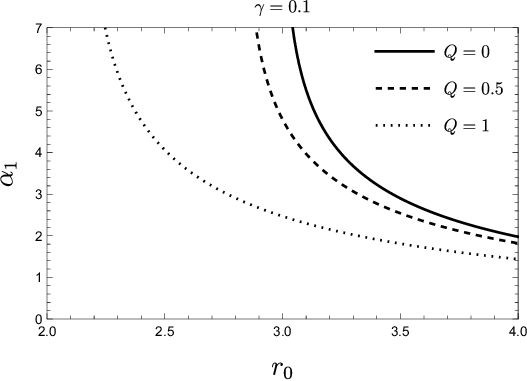}}
    \caption{The deflection angles $\alpha_1$ corresponding to  $g^{{\rm eff}(1)}_{\mu\nu}$  as a function of the  distance of closest approach, $r_0$,  for different values of the ModMax BH charge $Q$ and the nonlinear parameter $\gamma$ are shown.   $\alpha_2$ is larger than $\alpha_1$, $\alpha_2 = e^{ \gamma} \alpha_1$. The case $\gamma=0$ corresponds to RN BH and $Q=0$ corresponds to  Schwarzschild's deflection angle. }
    \label{fig:Fig5}
\end{figure}

\begin{figure}[H]
    \centering
    \includegraphics[width=0.45\textwidth]{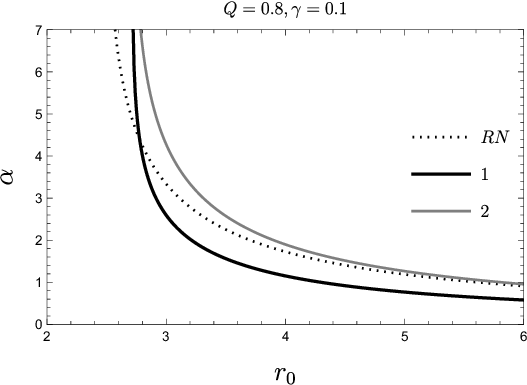}
    \caption{The deflection angles $\alpha_1$ and $\alpha_2$  for the two effective metrics and the corresponding to the RN-BH are shown. The relative magnitudes of  the deflection angles for  $r_0 > 3M$ is  $\alpha_1 < \alpha_{RN} < \alpha_2$ and $\alpha_1 = e^{ - \gamma} \alpha_2$.   }
    \label{fig:Fig6}
\end{figure}

In Fig. \ref{fig:Fig5} are shown the deflection angles produced by effective metric $g^{{\rm eff}(1)}_{\mu\nu}$ of the ModMax BH,  varying the BH charge and nonlinear parameter $\gamma$, resulting of the numerical integration.
For fixed $Q$ and $r_{0}$, increasing $\gamma$ diminishes the deflection angle $\alpha$
and the effect is the same, for fixed   $\gamma$ and $r_{0}$, increasing  the BH charge. The expression for $I_{2}(r_0)$  has the same form as for RN but with a charge screened by a factor  $e^{- \gamma}$. In Fig. \ref{fig:Fig6} we compare the deflection angle $\alpha_i, i=1,2$ of the two effective metrics with the RN one. 
When $r_{0} \rightarrow r_{c}$, the deflection angle reaches values greater than $2\pi$, in this case, the light ray turns around several times before either escaping the photosphere region or falling into the BH \cite{Eiroa2002}.

\subsubsection{The weak field limit of the deflection angle}

 The weak field limit deals with small deflection angles that can be determined through the Gaussian curvature and using the Gauss-Bonnet theorem with the optical metric method (details can be consulted in \cite{Gibbons2008, Okyay2022}).  

 The Gauss-Bonnet theorem connects the differential geometry of a surface with its topology; and the expression for a slight variation of  the  deflection angle $\delta \alpha$ is given by

\begin{equation}
 \delta \alpha=- \int \int_{D} K dS.
\end{equation}
where $dS=\sqrt{-g}dr d\phi$, $g$ is the determinant of the surface metric and $K$ is the Gaussian curvature; the range of integration is $r: \frac{b}{\sin \phi}<r<\infty$ and $\phi: 0<\phi<\pi$, with $b$ being the impact parameter.
The Gaussian curvature is $K=R/2$, being $R$ the Ricci scalar, and for the ModMax BH is
\begin{equation}
    K=\frac{2 M}{r^3}-\frac{3 M^2+3 e^{-\gamma} Q^2}{r^4}+\frac{6 e^{-\gamma } M Q^2}{r^5}-\frac{2 e^{-2 \gamma } Q^4}{r^6}.
\end{equation}

The determinants of the effective metrics,  ${\rm det}(g_{{\rm eff}(a)}^{\mu\nu})=g_{a}$, are 

\begin{equation}
   g_{1}=-\frac{e^{-\gamma } \left(Q_{e}^2+Q_{m}^2\right)^3}{\left[e^{\gamma } Q_{e}^2+e^{-\gamma}Q_{m}^2\right]^3} r^2, \qquad g_{2}=-\frac{ \left(Q_e^2+Q_m^2\right){}^6}{\left[2 \cosh (2 \gamma ) Q_e^2 Q_m^2+Q_e^4+Q_m^4\right]^3} r^2.
\end{equation}

\begin{figure}[H]
    \centering
    \includegraphics[width=0.5\textwidth]{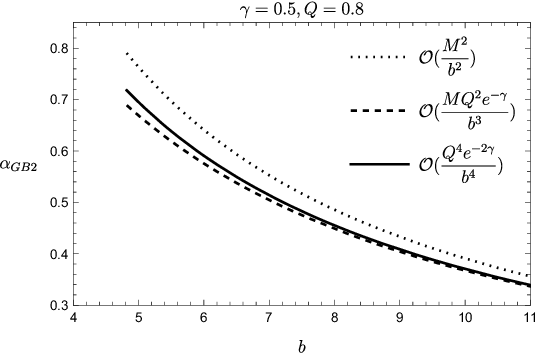}
    \caption{We compare the deflection angle $\alpha^{GB}_{2}$ 
    of Eq. (\ref{deflectionangle}) to different orders of $\mathcal{O}(\frac{M Q^2}{b})$.  The dotted line corresponds to the deflection angle considering the first two terms of Eq. (\ref{deflectionangle}), this is the deflection angle presented in \cite{Pantig2022}. The deflection angles taking the third and fourth terms in Eq. (\ref{deflectionangle}) are the dashed and solid lines,  respectively. }
    \label{fig:Fig7}
\end{figure}

Such that if we consider only electric charge $Q=Q_{e}$ ($Q_m=0$), the determinants are

\begin{equation}
    g_{1}=-e^{-4 \gamma } r^2, \qquad g_{2}=-r^2,
\end{equation}
and  the deflection angles  $\alpha_1^{GB}$ and $\alpha_2^{GB}$  are given by

\begin{equation}
    \alpha^{GB}_{2}= \left(\frac{4 M}{b}-\frac{3 \pi  e^{-\gamma } Q_e^2}{4 b^2}-\frac{3 \pi  M^2}{4 b^2}+\frac{8 e^{-\gamma } M Q_e^2}{3 b^3} -\frac{3 \pi  e^{-2 \gamma } Q_e^4}{16 b^4} \right) = e^{2 \gamma } \alpha^{GB}_{1},
    \label{deflectionangle}
\end{equation}
 the deflection angle $\alpha_2^{GB}$ only differs from the RN deflection in the screening of the charge.
In  \cite{Pantig2022} were considered only the first two terms in Eq. (\ref{deflectionangle}) for the deflection angle $\alpha^{GB}_{2}$. The effect introduced in the deflection angle by the third and fourth terms is illustrated in Fig. \ref{fig:Fig7}. Note that in \cite{Pantig2022} the deflection angle is overestimated.

\begin{figure}[H]
    \centering
    \includegraphics[width=0.5\textwidth]{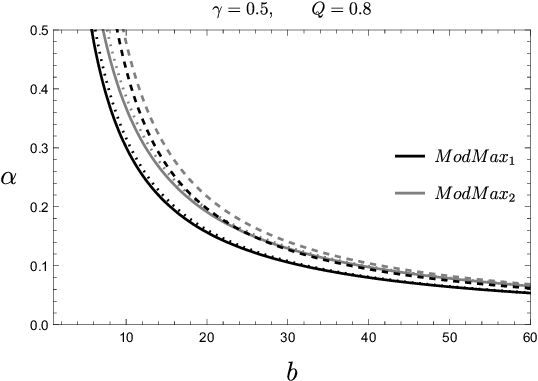}
    \caption{The plot compares the numerical result for $\alpha_1$ and $\alpha_2$ (dashed lines) with the Gauss-Bonnet calculation of the deflection angle in Eq. (\ref{deflectionangle}), $\alpha^{GB}_1$ and $\alpha^{GB}_2$ (solid lines), and the deflection angle presented in \cite{Pantig2022} (dotted lines). Note that as $b\rightarrow \infty$ the deflection angles approach the same limit.}
    \label{fig:Fig8}
\end{figure}

In Fig. \ref{fig:Fig8} we compare the numerical result for the deflection angle (dashed) with the weak deflection angles of the  ModMax BH
calculated according to the Gauss-Bonnet theorem for different orders of $\mathcal{O}(\frac{M Q^2}{b})$ (solid curves). The dotted line is the deflection angle calculated in \cite{Pantig2022} for the massless test particles. The deflection angles for the effective metric $g^{{\rm eff}(1)}$ are the black curves while the ones for the metric $g^{{\rm eff}(2)}$ are the gray curves. The lower value for the deflection angle is reached by considering all the terms in Eq. (\ref{deflectionangle}) for the metric $g^{{\rm eff}(1)}$, and the higher value corresponds to the numerical result for the effective metric $g^{{\rm eff}(2)}$.

\subsection{Redshift}
\label{sect:Redshift}

The gravitational redshift $z$ is a decrease in the frequency and photon energy, between the emitter and receiver points, due to the radiation traveling through a gravitational field   \cite{Salim2004, Payandeh2013, GuzmanBreton2021}, given by
\begin{equation}
    1+z=\frac{\omega_e}{\omega_{o}},
    \label{redshift}
\end{equation}
 where $\omega$ is the frequency and the subscripts $e$ and $o$ refer to the emitter and observer of the light pulse.
In general, the frequency of a photon measured by an observer with proper 4-velocity $U^{\mu}$ is given by

\begin{equation}
    \omega_{i}= -\left(k_{\mu} U^{\mu} \right)_{i},
\end{equation}
where  ``$i$" refers to ``$e$"   the emitter or ``$o$"  the observer, and $k_{\mu}=\{ \omega,  \vec{k} \}$ is the wave vector.

 We shall determine the redshift $z$ of a light pulse emitted from a particle moving in a circular orbit around the ModMax BH \cite{Herrera2015}. It is important to distinguish the trajectory of the emitter particle, which are timelike geodesics of the background metric  $g_{\mu\nu}$; while the photon trajectories are the null geodesics of the effective or optical metric $g^{{\rm eff}(a)}_{\mu\nu}$.

 Considering equatorial circular orbits, for the emitter particle, the  
 4-velocity $U^{\mu}$ is of the form   $U^{\mu}= ( U^{t}, U^{r}=0, U^{\theta}=0, U^{\phi})$,  in this case the redshift is given by 

\begin{equation}
    1+z=\frac{\omega_{e}}{\omega_{o}}=\frac{\left(k_{t}U^{t}+k_{\phi}U^{\phi} \right)_{e}}{\left(k_{t}U^{t}+k_{\phi}U^{\phi} \right)_{o}}.
    \label{redshiftmass}
\end{equation}

 From an analysis analogous to the one in Sec. \ref{sect:NullGeodesics}, now for the background geometry $g_{\mu \nu}$, we can determine the emitter particle velocity in an equatorial circular orbit;  its 
 energy, $E_{m}$,  and angular momentum, $L_{m}$, are  conserved quantities,  $U^t =-\frac{E_{m}}{g_{tt}}$ and $U^{\phi}=\frac{L_{m}}{g_{\phi\phi}}$.
 To determine $E_{m}$ and $L_{m}$   in terms of the BH parameters we use the circular orbit conditions (\ref{circ_orb_conds}),
where  the effective potential is given in  Eq. (\ref{effpotential}) making $g^{eff}_{\mu \nu} \mapsto g_{\mu \nu}$,  with $\delta= -1$, 
\begin{equation}
  V_{\rm eff, m} =f(r) \left(\frac{1}{r^2}-\frac{1}{f(r)}\frac{E_{m}^2}{L_{m}^2}+1\right),
    \label{effpotentialm}
\end{equation}
obtaining for $E_{m}$ and $L_{m}$, 
\begin{equation}
    E_{m}^2=\frac{f^2 (r)}{h(r)},\qquad L_{m}^2=\frac{\left({M}{r} - {e^{-\gamma}Q^2}\right)}{h(r)},
\end{equation}
where we defined $h(r)=1-3M/r+2e^{-\gamma}Q^2/r^2$. Writing the angular velocities of the emitter particle as $\Omega_{i}=U^{\phi}_{i}/U^{t}_{i}$, 
\begin{equation}
    \Omega_{i}^2=\frac{M r_{i}-e^{-\gamma}Q^2}{r_{i}^4}, \qquad i=e,o,
\end{equation}
and the frequency $\omega_{i}$ of a photon  is given by 

\begin{equation}
    \omega_{i}=-U_{i}^{t}\left(k_{t}+k_{\phi}\Omega\right)_{i}.
\end{equation}

The photons move along null geodesics of the effective metric $g_{{\rm eff}(a)}^{\mu\nu}k_{\mu}k_{\nu}=0$, where $k^t = E/f(r)$ and $k^\phi =L/r^2$; then in terms of the impact parameter, $b=L/E$, \cite{Gravitation}, then the expression for the photon frequency is 

\begin{equation}
    \omega_{i}=-\sqrt{\frac{1}{h(r_{i})}}\left(\frac{g^{{\rm eff}(a)}_{tt}}{f(r)}+b \frac{g^{{\rm eff}(a)}_{\phi\phi}}{r^4}\sqrt{{M r-e^{-\gamma}Q^2}}\right)_{i},
    \label{omegaib}
\end{equation}
and considering the critical impact parameter in Eq. (\ref{critimpactparam})

\begin{equation}
    \omega_{i}=-\sqrt{\frac{1}{h(r_{i}) }}\left\{ 
 \frac{g^{{\rm eff}(a)}_{tt}}{f(r)} \pm \sqrt{\frac{- g_{tt}^{{\rm eff}(a)}}{f(r)}\frac{g^{{\rm eff}(a)}_{\phi\phi}}{r^2}} \sqrt{\frac{M r-e^{-\gamma}Q^2}{f(r)r^2}}\right\}_{i},
\label{wi} 
\end{equation}
where the $\pm$ sign refers to a receding and an approaching emitter particle, respectively. Such that the blueshift ($+$) /redshift ($-$), we defined $A^2(r)=(M r-e^{- \gamma}Q^2)/(f(r)r^2)$. For the effective metrics $g_{{\rm eff}(a)}, a=1,2$ are

\begin{equation}
    (z_{1})_{\pm} =\left( \frac{h(r_{o})}{h(r_{e})}\right)^{1/2} 
    \left( \frac{-e^{-\gamma } \pm A(r_{e}) }{-e^{-\gamma } \pm A(r_{o})}\right)-1
    \label{redshiftroinbh1}
\end{equation}

\begin{equation}
    (z_{2})_{\pm} =\left( \frac{h(r_{o})}{h(r_{e})}\right)^{1/2}  \left(\frac{-e^{-\gamma } \pm e^{\gamma } A(r_{e})}{-e^{-\gamma } \pm e^{\gamma }A(r_{o})}\right)-1,
\end{equation}
Considering an observer at infinity, $r_{o}\rightarrow \infty$, then $A(r_{o})=0, \quad h(r_{o})=1$, and the redshifts are

\begin{equation}
    (z_{1})_{\pm} =-\sqrt{\frac{1}{h(r_{e})}}\left[-1 \pm e^{\gamma }A(r_{e})\right]-1
\end{equation}

\begin{equation}
   (z_{2})_{\pm} =-\sqrt{\frac{1}{h(r_{e})}}\left[ -1 \pm e^{2\gamma }A(r_{o})\right] -1
\end{equation}

\begin{figure}[H]
    \centering
    \subfigure{\includegraphics[width=0.45\textwidth]{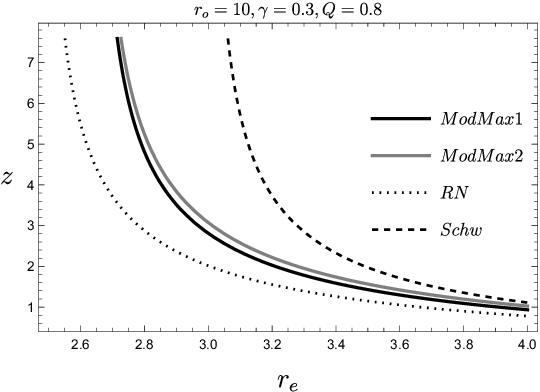}}
    \subfigure{\includegraphics[width=0.45\textwidth]{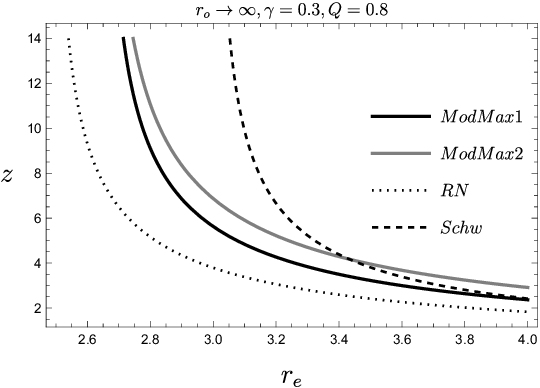}}
    \caption{The redshifts of a pulse emitted from a particle moving in a circular orbit around the Schwarzschild, RN, and ModMax BHs as a function of the emitter position $r_{e}$ measured by an observer in the vicinity of the  BH ($r_{o}=10$) are illustrated to the left. The case where the observer is located at infinity ($r_{o}\mapsto \infty $) is illustrated to the right. For a given value of $r_{e}$, the redshift for an observer in the vicinity of the BH is smaller than the one measured by the observer at infinity; in terms of the observed frequency   $\omega_{o}$, the one measured by the observer in the vicinity is higher than the frequency measured by the observer located at infinity.  In this plot  $\gamma=0.3$ and   $Q=0.8$}
    \label{fig:Fig9}
\end{figure}

\begin{figure}[H]
    \centering
    \subfigure{\includegraphics[width=0.45\textwidth]{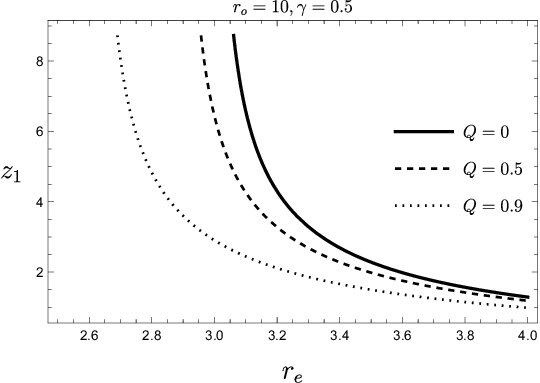}}
    \subfigure{\includegraphics[width=0.45\textwidth]{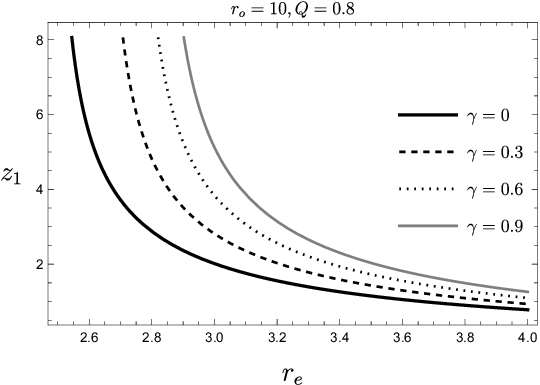}}    \caption{We plot the redshifts produced by $(g_{{\rm eff}(1)}^{\mu\nu})$  as a function of the emitter position $r_{e}$, Eq. (\ref{redshiftroinbh1}); to the left for different values of the BH charge with a fixed $\gamma=0.5$; $Q=0$ corresponds to Schwarzschild BH.  And to the right for different values of the nonlinear parameter $\gamma$  with fixed charge $Q=0.8$; $\gamma=0$ corresponds to RN BH. }
    \label{fig:Fig10}
\end{figure}

In Fig. \ref{fig:Fig9} are plotted the redshifts as a function of the emitter position, $r_e$, as measured at $r_o$ by an observer in the vicinity of the BH and an observer at infinity. Note the difference in the z-range. In terms of the frequency, for a given position of the emitter $r_{e}$, the frequency measured by an observer in the vicinity of the BH is larger than the frequency measured by the observer located at infinity. 
In Fig. \ref{fig:Fig10} is illustrated the effect on the redshift, Eq. (\ref{redshiftroinbh1}),  of varying the charge and NLED parameter of the ModMax  BH, for the effective metric $(g_{{\rm eff}(1)}^{\mu\nu})$, for an observer located in the vicinity of the  BH.  For a fixed  $\gamma$ the increase of the BH charge diminishes the redshift (increases the frequency); and with a fixed BH charge $Q$,  increasing  $\gamma$  increases the redshift   (diminishes the frequency).
In  the case  that both the emitter and observer are static and the observer's position tends to infinity, $r_{o}\rightarrow \infty$, then $U^\phi =0$, and  from $U^{\mu}U_{\mu}=-1$ we obtain 

\begin{equation}
U^t=\sqrt{-\frac{1}{g_{tt}}};    
\end{equation}
then the  frequency $\omega_{i}$ of a light pulse is

\begin{equation}
    \omega_{i}=-\left( k_{t}U^t \right)_{i}=\left(- \frac{g_{tt}^{{\rm eff}(a)}}{f(r)}\sqrt{-\frac{1}{g_{tt}}} \right)_{i};
\end{equation}
while for the observer at infinity  $r_{o}\rightarrow \infty$,  $\omega_{o}=1$, and the redshift is
\begin{equation}
    1+z=-\frac{g_{tt}^{{\rm eff}(a)}}{f(r)}\sqrt{\frac{1}{f(r)}} .
\end{equation}

\begin{figure}[H]
    \centering
    \includegraphics{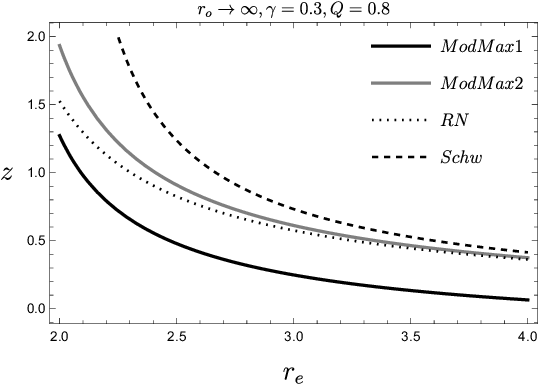}
    \caption{The redshifts of a light pulse emitted from a static particle at $r_{e}$ in the vicinity of Schwarzschild, RN, and ModMax  BHs,  measured by an observer located at infinity ($r_o \mapsto \infty$).  The frequency corresponding to the effective metric $g_{\mu\nu}^{{\rm eff}(1)}$ is higher than the one for the effective metric $g_{\mu\nu}^{{\rm eff}(2)}$, since $z_{1}<z^{\rm RN}<z_{2}<z^{\rm Schw}$. In this plot are fixed  $\gamma=0.3$ and  $Q=0.8$.}
    \label{fig:Fig11}
\end{figure}

The redshifts corresponding to the two effective metrics are given by

\begin{equation}
    z_{1}=\frac{e^{-\gamma } \left(e^{-\gamma } Q_{e}^2+e^{\gamma } Q_{m}^2\right)}{Q_{e}^2+Q_{m}^2} \sqrt{\frac{1}{f(r)}}-1,
\end{equation}

\begin{equation}
    z_{2}=\sqrt{\frac{1}{f(r)}}-1;
\end{equation}
$z_{2}$ corresponds to the redshift of a massless particle in a RN BH with a screened charge $Q^2 \mapsto e^{- \gamma} Q^2$. In Fig. \ref{fig:Fig11} are compared the redshifts for the ModMax, the RN, and the Schwarzschild BHs as a function of the emitter position, for a static observer ($r_o \mapsto \infty$);   the frequency of the light pulse moving in the effective metric $g_{\mu\nu}^{{\rm eff}(1)}$ is larger than the one of the light pulse moving in $g_{\mu\nu}^{{\rm eff}(2)}$ or in the RN BH and Schwarzschild BH metric. The relative magnitudes of the redshift, $z_{1}<z^{\rm RN}<z_{2}<z^{\rm Schw}$, tells us that the ModMax BH creates a weaker gravitational field (for the photons) than RN BH since the photon in the vicinity of the ModMax BH suffers a smaller loss of energy to climb the gravitational potential than the RN one.

\begin{figure}[H]
    \centering
    \includegraphics[width=0.7\textwidth]{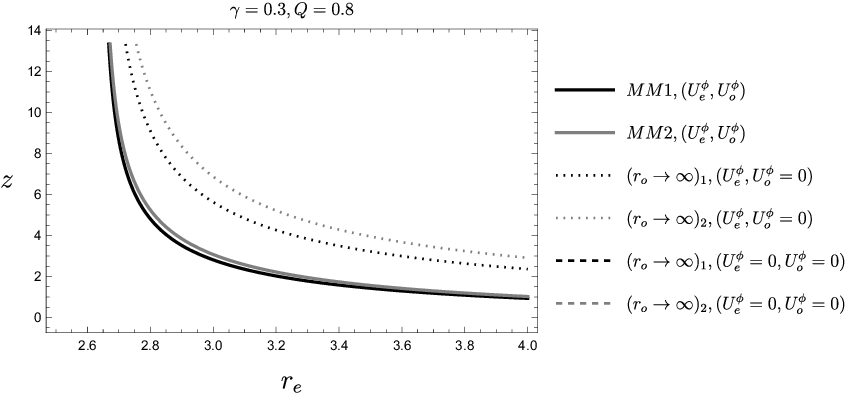}
    \caption{We plot the redshift of a light pulse in three situations: (1) the emitter is a particle moving in a circular orbit around the ModMax BH, and the observer is in the vicinity of the BH orbiting as well (solid curves), $U^{\phi}_{e} \ne 0$ and $U^{\phi}_{o} \ne 0$; (2) the emitter is orbiting around the ModMax BH and the observer is located at infinity (dotted), i.e. $U^{\phi}_{e} \ne 0$ and $U^{\phi}_{o} = 0$. (3) Both, emitter and observer,  are static at infinity (dashed curves) $U^{\phi}_{e} = 0$ and $U^{\phi}_{o} = 0$.  BH parameters are fixed:  $\gamma=0.3$, $Q=0.8.$}
    \label{fig:Fig12}
\end{figure}

In Fig. \ref{fig:Fig12}, we compare the redshifts of a light pulse moving in the effective metrics $g_{\mu\nu}^{{\rm eff}(a)}, a=1,2$, emitted from a particle orbiting the BH, for different positions of the observer. The larger frequency (smaller $z$) would be measured by the observer at infinity when the emitter is static and the pulse of light moves according to the effective metric $g_{\mu\nu}^{{\rm eff}(1)}$.
\subsubsection{The kinematic redshift}

The kinematic redshift is relevant in relation to astrophysical observations, for instance in measurements of galaxies redshift \cite{Herrera2015}. It is given by the difference between the redshift in Eq. (\ref{redshiftmass}) and the redshift of a photon emitted from  $b=0$, using Eq. (\ref{omegaib})

\begin{equation}
z_{k}=\frac{\omega_{e}}{\omega_{o}}-\left(\frac{\omega_{e}}{\omega_{o}}\right)_{b=0}= b \frac{U^{t}_{e}}{U^{t}_{o}}\frac{\Omega_{o}-\Omega_{e} }{\left(1-b \Omega_{o} \right)}.
\end{equation}

Considering the two effective metrics, the kinematic redshifts $z_{k_{i}}$ are given by

\begin{equation}
  z_{k1}=e^{\gamma}
  \sqrt{\frac{h(r_{o})}{h(r_{e})}}
  \left(
  \frac{A(r_{e})-A(r_{o})}{\mp 1+e^{\gamma}A(r_{o})} \right),
\end{equation}

\begin{equation}
  z_{k2}=e^{\gamma }\sqrt{\frac{h(r_{o})}{h(r_{e})}} \left( \frac{A(r_{e})-A(r_{o})}{\mp e^{-\gamma }+e^{\gamma }A(r_{o})} \right).
\end{equation}

\begin{figure}[H]
    \centering
    \subfigure{\includegraphics[width=0.45\textwidth]{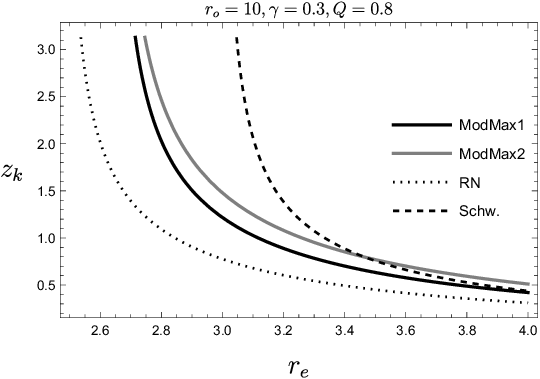}}
    \subfigure{\includegraphics[width=0.45\textwidth]{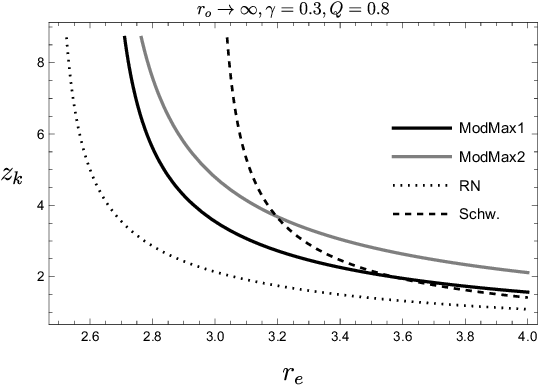}}
    \caption{ The kinematic redshifts  $z_{k}$ for Schwarzschild, RN, and ModMax BHs are illustrated as a function of the emitter position; to the left for an observer in the vicinity of the BH  and to the right for an observer at infinity. For a given $r_{e}$ the redshift measured by an observer at infinity is larger (smaller frequency) than the one for the observer in the vicinity of the BH.}
    \label{fig:Fig13}
\end{figure}

In the case  that the observer is located at infinity with respect to the center of the BH, $r_{\rm o} \rightarrow \infty$,  the kinematic redshifts are 

\begin{equation}
  z_{k1}=\mp e^{\gamma }\sqrt{\frac{A^2(r_{e})}{h(r_{e})}}, 
\end{equation}
and
\begin{equation}
  z_{k2}=\mp e^{2\gamma }\sqrt{\frac{A^2(r_{e})}{h(r_{e})}}.  
\end{equation}
The kinematic redshift is illustrated in Fig. \ref{fig:Fig13} for the Schwarzschild, RN, and ModMax BHs, as observed near the BH and at infinity; the relation between the redshifts of the two ModMax BH effective metrics is $z_{k2}= e^{\gamma} z_{k1}$.
Note that there is a difference in the range between the kinematic redshift and the redshift in Fig. \ref{fig:Fig9};  the kinematic redshift reaches smaller values for a given position of the emitter.

\subsection{Shadow}
\label{sect:shadow}

In the vicinity of a BH, there is the possibility of unstable circular orbits of massless particles and photons; any perturbation will cause either the particle to fall into the BH or to escape to infinity. Photons that escape to infinity produce a shadow cast perceived by an external observer. In \cite{Pantig2022} is presented the shadow produced by massless particles calculated considering the null geodesics of the background metric $g_{\mu\nu}$, and it is analyzed the effect of the nonlinear parameter $\gamma$ on the shadow measured by an observer co-moving with the cosmological expansion. As we have shown, there are two possible photon paths, corresponding to the null geodesics of the two effective or optical metrics. One of the trajectories corresponds to the null geodesics of the background metric, which we have denoted by  $g_{\mu\nu}^{{\rm eff}(2)}= g_{\mu\nu} $, and is the one studied in \cite{Pantig2022}. 

The second possible trajectory, the null geodesics of   $g_{\mu\nu}^{{\rm eff}(1)}$ generate a second shadow and a second absorption cross section (ACS), that we present in the following.

To calculate the radius of the shadow for a distant observer in a SSS spacetime, we can use the optical approximation where the critical impact parameter corresponds to the radius of the shadow $r_{sh}=b_{c}$, \cite{Perlick2022, Vagnozzi2023}, then from Eq. (\ref{impactparam12}),

\begin{equation*}
    r_{sh1}=\sqrt{\frac{e^{-2\gamma}r_{c}^2}{f(r_{c})}}, \qquad r_{sh2}=\sqrt{\frac{r_{c}^2}{f(r_{c})}};
\end{equation*}
recall that $r_c$ is the radius of the photosphere, i.e. the radius of the unstable circular orbits (UCO) of the photon, that for a NLED BH has to be calculated from the effective metric.
In order to compare the ModMax BH shadow with the observed angular radius Sgr $A^{*}$ we follow the methodology used in \cite{Vagnozzi2023} considering the uncertainty allowed by the EHT observations, with $\delta$ as the fractional deviation between the shadow radius $r_{sh}$ and the shadow radius of a Schwarzschild BH $\delta=r_{sh}/(3M\sqrt{3})-1$. Assuming Gaussianity, the values for $\delta$ following $1\sigma $ and $2\sigma$ intervals are $-0.125 \gtrsim \delta \gtrsim 0.005$ for $(1\sigma)$ and $ -0.19 \gtrsim \delta \gtrsim 0.07 $ for $(2\sigma)$.

\begin{figure}[H]
    \centering
    \includegraphics{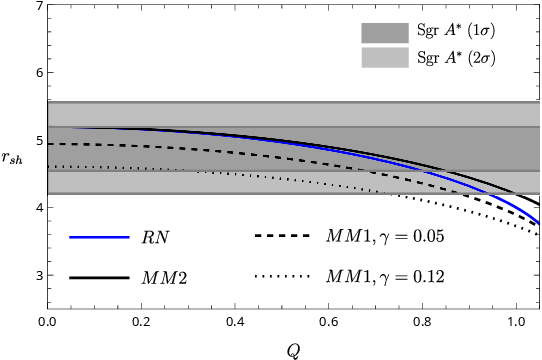}
    \caption{The shadow radius generated by the two effective metrics of the ModMax BH as a function of the BH charge.  For the effective metric $g^{{\rm eff}(1)}$ (dashed and dotted) we fixed $\gamma=0.05$ and $\gamma=0.12$; while for the effective metric $g^{{\rm eff}(2)}$, $\gamma=0.12$. The shadow radius for the RN BH (blue line) is the reference to define the effects due to the electromagnetic nonlinearity.  The dark gray and light gray bands correspond to the $1\sigma$ and $2\sigma$, respectively, for the Sgr $A^{*}$ BH; the line at $r_{sh}= 3 \sqrt{3}  \approx 5.19$ corresponds to the Schwarzschild's shadow.   } 
    \label{fig:Fig14}
\end{figure}

In Fig. \ref{fig:Fig14} is shown the shadow radius generated by the two effective metrics and the corresponding to the RN BH as a function of the BH charge $Q$, contrasted by the EHT for Sgr $A^{*}$. As the charge increases the radius of the shadow diminishes and the restrictions imposed by the $1\sigma$ allow us to 
set bounds in the BH charge or in $\gamma$.  For $g^{{\rm eff}(1)}$ we are considering different values of $\gamma$, and note that as $\gamma$ increases the shadow radius is no longer in the $1\sigma$ region. Consistency with the observations of the shadow of Sgr $A^{*}$ for RN BH restricts the BH charge to $Q \le 0.8 M$; for the ModMax BH this constraint becomes $Q e^{- \gamma /2} \le 0.8 M$, or
$\gamma \ge -2 \ln ( 0.8 M/Q)$. For instance, if we set the value of $\gamma=0.12$ as a lower bound of $\gamma$, the radius of the shadow is no longer in the $1\sigma$ region for values of the charge $Q\ge 0.2991$. For lower values of $\gamma$, a greater range of $Q$ is allowed, i.e. the constraints for $\gamma$ depend on the values of $Q$.

\subsubsection{The shadow for an observer at infinity}
Alternatively,  the shadow can be determined by considering a light ray from the observer to the UCO  at an angle $\psi_{sh}$ with respect to the radial direction. The expression for the deviation angle associated with the shadow is  \cite{Perlick2015}

\begin{equation}
    \sin^2(\psi_{sh})_{a}=-(b^2_{ca})\left(\frac{g^{{\rm eff}(a)}_{\phi\phi}}{g^{{\rm eff}(a)}_{tt}}\frac{f^2(r_{o})}{r^{4}_{o}} \right)_{r_{o}}.
\end{equation}

In terms of the two impact parameters we obtain
\begin{equation}
    \sin^2(\psi_{sh})_{1}=e^{2\gamma}b^2_{c1}\frac{f(r_{o})}{r^2_{o}}, \qquad \sin^2(\psi_{sh})_{2}=b^2_{c2}\frac{f(r_{o})}{r^2_{o}}.
    \label{shadowangle}
\end{equation}
The radius of the BH shadow is approximated in terms of the observer's position and the  angle $\psi_{sh}$ as \cite{Amaro2022, Amaro2020, Perlick2018}
\begin{equation}
    r_{\rm sh}= r_{o}\tan{\psi_{sh}} \approx r_{o}\sin{\psi_{sh}}.
    \label{shadowradius}
\end{equation}
Then using Eq. (\ref{impactparam12}) we can calculate the radius of the shadow,  assuming that the observer is located at infinity  $r_{o} \mapsto \infty$. Despite the existence of two values for the impact parameter, corresponding to the two effective metrics, in this approximation,  the static observer at infinity will detect only one shadow of the ModMax BH; the reason is that there is a cancellation of the screening factor, using the relation  $e^{2\gamma}b_{c1}^2=b_{c2}^2$   in Eqs. (\ref{shadowangle}) and the approximation in Eq. ( \ref{shadowradius}) implies that  $\sin^2(\psi_{sh})_{1}=\sin^2(\psi_{sh})_{2}$.

\begin{figure}[H]
    \centering
   \includegraphics[width=0.45\textwidth]{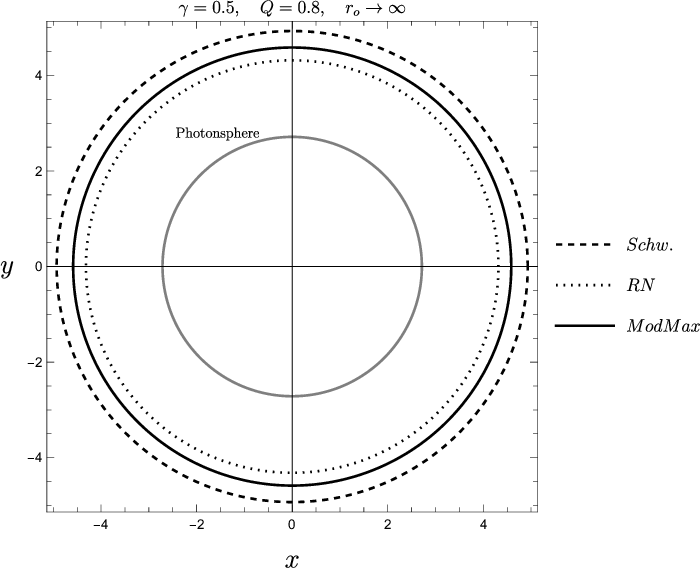}
    \caption{The radii of the BH shadows for Schwarzschild, RN, and the ModMax BHs are illustrated considering that the position of the observer is at infinity, $r_{o}\rightarrow \infty$; the relative sizes of the shadow radii are $  r_{\rm sh}^{RN} <   r_{\rm sh}^{MM} <   r_{\rm sh}^{Schw}$.  The gray circumference is the photosphere $r_{c}$ of the ModMax BH. In this plot $Q=0.8$   and  $\gamma=0.5$ and  polar coordinates  are  $x=r_{sh}\cos(\zeta)$, $y=r_{sh}\sin(\zeta)$. }
    \label{fig:Fig15}
\end{figure}

In Fig. \ref{fig:Fig15} are illustrated the radii of the BH shadows for Schwarzschild, RN, and the ModMax BHs, considering that the position of the observer is at infinity,  $r_{o}\rightarrow \infty$.  Notice that the shadow radius of the ModMax BH is larger than the  RN  one and smaller than Schwarzschild's, $  r_{\rm sh}^{RN} <   r_{\rm sh}^{MM} <   r_{\rm sh}^{Schw}$, i.e.
the radius of the shadow for the effective metric $g^{{\rm eff}(2)}$ is located between the radius of the shadow for RN BH and the one for the Schwarszchild BH in agreement with the results in \cite{Pantig2022}.

\begin{figure}[H]
    \centering
    \includegraphics{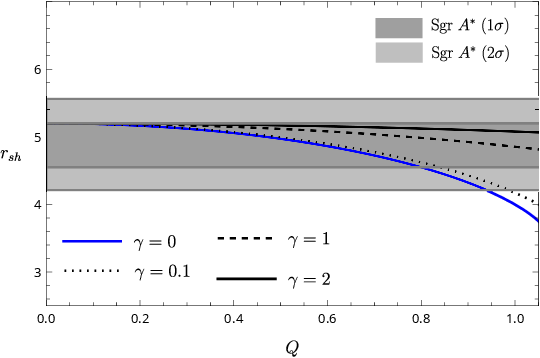}
    \caption{The shadow radius of the ModMax BH (black lines) as a function of the BH charge, for different values of $\gamma$, is compared to the shadow radius of the RN BH (black line). The dark gray and light gray bands correspond to the $1\sigma$ and $2\sigma$, respectively, for the Sgr $A^{*}$ BH.}
    \label{fig:Fig16}
\end{figure}

 In Fig. \ref{fig:Fig16} is shown the shadow radius for the ModMax BH, for different values of $\gamma$, as well as the RN's, as a function of the charge $Q$, contrasted to the constraints by the EHT for Sgr $A^{*}$. As the charge increases the radius of the shadow decreases. In the RN BH case constraints on the value of the charge can be deduced. In the case of the ModMax BH there is the additional parameter $\gamma$ and the radius of the shadow depends on both $(\gamma,Q)$. The screening of $e^{- \gamma}$ allows larger values for the charge; such that for the range $0< \gamma < 3.6$ the ModMax BH shadow remains in the $(1 \sigma)$ interval. 

   For $\gamma=3.6$ the shadow radius is $r_{sh}=5.19521$ with a BH charge of $Q=0.2$; while  $r_{sh}=5.18096$ with a BH charge $Q=0.8$; both tend to the shadow radius for the Schwarszchild BH, $r_{sh, SHW}=5.19615$ within the uncertainty established by the $(1 \sigma)$ interval. 
\subsubsection{Absorption cross section }   

\begin{figure}[H]
    \centering
   \includegraphics[width=0.45\textwidth]{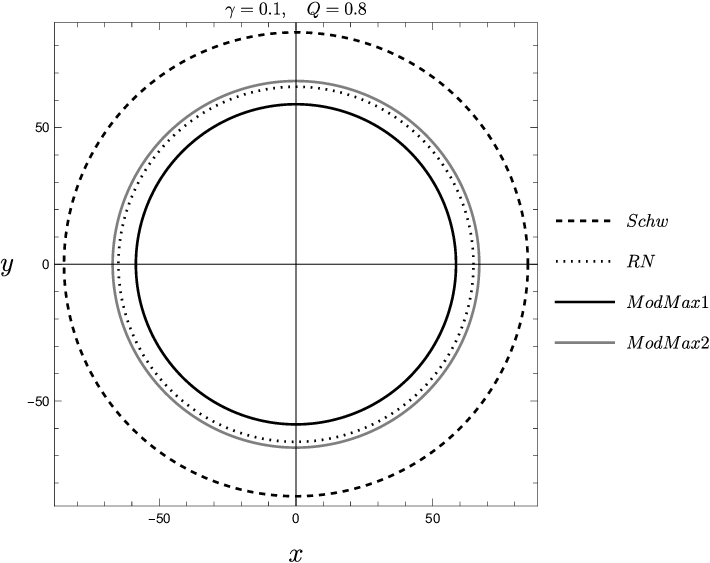}
    \caption{ We plot the absorption cross sections (ACS)  of the Shwarzschild, RN, and the two effective metrics of the ModMax BH. The ACS for the effective metric $g^{{\rm eff}(1)}$ is the black circle (the smallest), and for the effective metric $g^{{\rm eff}(2)}$ is the gray one. The ACS for the effective metric $g^{{\rm eff}(2)}$ is larger than the RN one.  The larger ACS corresponds to the Schwarszchild BH. This plot is in polar coordinates, with $x=r_{sh}\cos(\zeta)$, $y=r_{sh}\sin(\zeta)$. }
    \label{fig:Fig17}
\end{figure}

On the plane of a distant observer the boundary of the BH shadow marks the apparent image of the photon region that separates capture orbits from scattering orbits.
The absorption cross section (ACS) originates in the photons moving in UCO that are captured by the BH; in the limit of geometrical optics \cite{Crispino2008} the ACS is given by $\sigma_{a}=\pi b_c^2$, with $b_c$ being the impact parameter of the UCO. For the ModMax BH, due to the existence of two impact parameters corresponding to the two effective metrics Eqs. (\ref{eff_metric_1}) and (\ref{eff_metric_2}), there are two ACS,

\begin{equation}
    \sigma_{a}=\pi b_{ca}^2, \quad a=1,2.
\end{equation}
that are shown in Fig. \ref{fig:Fig17}. The relation between the two ModMax BH ACS is $\sigma_2= e^{2 \gamma} \sigma_1$ and the relative magnitude compared with Schwarzschild and RN ACS is 
$\sigma^{Schw} > \sigma_2 > \sigma^{RN} > \sigma_1$,
in agreement with the shadow sizes.

\section{Conclusions}

In nonlinear electrodynamics, light propagates along null geodesics of the effective optical metric. Since there are two effective metrics for the ModMax nonlinear theory,  birefringence takes place. Coupling the ModMax nonlinear electrodynamics with the Einstein equations, SSS solutions with horizons have been found in \cite{Maceda2020}, characterized by the parameters of mass, electric or magnetic charges, and the ModMax parameter $\gamma$. It turns out that the ModMax BH has the form of the RN metric with a charge screened by the factor $e^{-\gamma/2}$. 
 To study light propagation in the vicinity of the ModMax BH as the background metric,    we determine the two effective metrics.  One of the consequences of the NLED conformal invariance is that
one of the effective metrics turns out to be the background metric, $g^{{\rm eff}(2)}_{\mu\nu}= g_{\mu\nu}$, therefore the studied effects
(phase velocities, light trajectories,  light ray deflection, redshift of light coming from the BH, shadow) are qualitatively the same as the ones for massless particles for a RN BH, but with a smaller charge, due to the screening produced by the nonlinear parameter $\gamma$, in the form of $Q^2 \mapsto e^{- \gamma} Q^2$. While the effects due to the second effective metric $g^{{\rm eff}(1)}_{\mu\nu}$ are more interesting and cannot be deduced from the background metric. 
  
Phase velocities of the two possible light trajectories in the vicinity of the ModMax BH, are always slower than in the neighborhood of the Reissner-Nordstrom BH. For a light pulse propagating in a purely radial direction, there is no birefringence; however, if the velocity has purely angular or radial and angular components then there appears birefringence.

 Regarding light trajectories in the vicinity of the ModMax BH, there are two possible unstable circular orbits (UCO); the one corresponding to the background metric exhibits a radius larger than the one of the RN BH and smaller than Schwarzschild's. Such that one of the ModMax BH light trajectories elapses between the ones corresponding to the RN and Schwarzschild BHs. The second UCO has a radius larger than the previously mentioned and therefore the distance to reach the horizon is larger.

 The deflection angles are determined numerically in the regime of strong deflection and then analytically considering the weak field limit, by means of the Gauss-Bonnet theorem. From the numerical approach, we find the deflections produced by the two effective metrics of the ModMax BH; for $g^{{\rm eff}(1)}_{\mu\nu}$ we find a smaller deflection angle by a factor of $e^{-\gamma}$ additional to the screening of the charge. The effect of the nonlinearity on the other effective metric $g^{{\rm eff}(2)}_{\mu\nu}$ is only the screening of the BH charge. In Fig. \ref{fig:Fig5} it is shown that for fixed $Q$ and closest distance, $r_{0}$, increasing the nonlinear parameter $\gamma$ diminishes the deflection angle $\alpha$ and for fixed $\gamma$ and $r_{0}$, increasing $Q$ also diminishes the deflection angle $\alpha$. In Fig. \ref{fig:Fig6} it is illustrated that for fixed  $Q$ and $\gamma$ the RN deflection angle lies between the ones for the two effective metrics.  In the weak field limit, we verify that for an observer located at infinity, the light deflection angle is the same for the two effective metrics.
The results for the weak deflection angle are in agreement with the ones presented in  \cite{Pantig2022}. Moreover we present a more accurate expression that includes three additional terms of the orders  $(M/b)^2$,  $MQ^2/b^3$ and $Q^4/b^4$, where $b$ is the impact parameter; we present graphically the effects introduced by the extra terms (see Fig. \ref{fig:Fig7}), resulting in a smaller deflection angle, showing that in \cite{Pantig2022} the deflection is overestimated.

The redshift measured by an observer in the vicinity of the BH and by an observer located at infinity is determined for an emitter that is orbiting around the ModMax BH and considering several positions of the observer. The frequency measured by an observer in the vicinity of the BH is larger than the frequency measured by the observer located at infinity.
We emphasized the difference between the emitter trajectory along a timelike geodesic of the background metric $g_{\mu\nu}$ and the photon motion along null geodesics of the effective metric $g^{{\rm eff}(a)}_{\mu\nu}$.  Due to the existence of two effective metrics, there are two possible redshifts.  The lower redshift corresponds to the effective metric  $g^{{\rm eff}(1)}_{\mu\nu}$ when the observer is at infinity and the emitter is static;  in terms of the frequency, the larger frequency would be measured by the observer at infinity when the emitter is static and the pulse of light moves according to the effective metric  $g^{{\rm eff}(1)}_{\mu\nu}$. In general the redshift produced by $g^{{\rm eff}(1)}_{\mu\nu}$ is smaller than the one for the RN BH, while the one in the metric $g^{{\rm eff}(2)}_{\mu\nu}$ is larger than the one for RN and smaller than the one of Schwarzschild BH. 

We also present the kinematic redshift, $z_{k}$ which is important in measurements of galaxy redshifts; $z_{k}$ takes into account the peculiar velocities. In general, the kinematic redshift $z_{k}$ is smaller than the previously described redshifts $z$.

The radii of the shadow for the effective metrics are analyzed.
The shadow corresponding to the effective metric  $g^{{\rm eff}(2)}$,  equivalent to the background metric,  was presented in \cite{Pantig2022}. Our analysis for the effective metric $g^{{\rm eff}(1)}_{\mu\nu}$ extends the study to take into account the second possible photon trajectory.
We set constraints of acceptable pairs of BH charge and nonlinear parameter 
$(Q, \gamma)$ that fall in the $1\sigma$ interval of the observations of the shadow of Sagittarius $A^{\ast}$. For RN BH the maximum acceptable BH charge is $Q=0.8$, due to the screening, with $\gamma=2$ BH charges of $Q \approx 1.1$ as allowed. For $\gamma > 3.6$ the ModMax BH shadow is indistinguishable from the Schwarzschild's shadow.

In the limit of geometrical optics, we determined the two possible absorption cross sections (ACS) of the ModMax BH,   due to the two values for the critical impact parameter;  the areas of the ACS are ordered in magnitude as  $\sigma_{1}<\sigma_{\rm RN}<\sigma_{2}<\sigma_{Schw}$.

Finally, we emphasized that while the massless particle effects of the ModMax BH  background metric (equivalently to $g^{{\rm eff}(2)}_{\mu\nu}$) can be  
described as a transition from the RN to the Schwarzschild BH, and approaching Schwarzschild as $\gamma$ increases, 
the results provided by the other effective metric ($g^{{\rm eff}(1)}_{\mu\nu}$ in our work) for the photon behavior cannot be deduced from the geometric or background metric of the electromagnetic nonlinear charged BH.

\vspace{0.5cm}
\textbf{Acknowledgments}: The work of E G-H has been sponsored by CONACYT-Mexico through the Ph. D.  scholarship No.761878.

\end{document}